\documentclass[twocolumn,pra,showpacs,amsmath,amssymb]{revtex4}
\usepackage{graphicx}
\usepackage{amssymb,amsthm,amsfonts,amstext}
\usepackage{enumerate}
\usepackage{url}
\def\be{\begin{equation}}
\def\ee{\end{equation}}
\def\bea{\begin{eqnarray}}
\def\eea{\end{eqnarray}}
\def\bma{\begin{mathletters}}
\def\ema{\end{mathletters}}

\def\0{\overline{0}}

\def\q0{\underline{0}}

\def\C{{\mathbb C}}
\def\id{{\mathbb I}}

\def\tr{\mbox{tr}}

\def\one{\leavevmode\hbox{\small1\normalsize\kern-.33em1}}

\def\bra#1{\langle#1|} \def\ket#1{|#1\rangle}
\def\braket#1#2{\langle#1|#2\rangle}

\def\proj#1{\ket{#1}\!\bra{#1}}

\def\id{{\mathbb I}}
\def\tr{\mbox{tr}}

\begin{document}

\title{The Physics of crypto-nonlocality}

\author{Miguel Navascu\'es$^{1,2}$}
\affiliation{$^1$H.H. Wills Physics Laboratory, University of Bristol, Tyndall Avenue, Bristol, BS8 1TL, United Kingdom\\$^2$F\'isica Te\`orica: Informaci\'o i Fen\`omens Qu\`antics,
Universitat Aut\`onoma de Barcelona, 08193 Bellaterra (Barcelona), Spain}

\begin{abstract}
In 2003, Leggett introduced his model of crypto-nonlocality based on considerations on the reality of the photon polarization. In this note, we prove that, contrary to hints in subsequent literature, crypto-nonlocality does not follow naturally from the postulate that polarization is a realistic variable. More explicitly, consider physical theories where: a) faster-than-light communication is impossible; b) all physical photon states have a definite polarization; and c) given two separate photons, if we measure one of them and post-select on the result, the measurement statistics of the remaining system correspond to a photon state. We show that the outcomes of any two-photon polarization experiment in these theories must follow the statistics generated by measuring a separable two-qubit quantum state. Consequently, in such experiments any instance of entanglement detection -and not necessarily a Leggett inequality violation- can be regarded as a refutation of this class of theories.
\end{abstract}

\maketitle

In 1935, Einstein, Podolski and Rosen (EPR) argued that quantum mechanics could not be considered a complete theory, as it did not assign definite values to the different observable quantities \cite{EPR}. This property was called \emph{realism}, and EPR believed that some future physical theory would combine this feature with the astounding predictive powers of quantum mechanics. Bell later reformulated the EPR argument \cite{bell}: by considering experiments conducted by two separate parties, he showed that the predictions of quantum mechanics were incompatible with any realistic theory where distant observables could not influence each other at superluminal speeds. Very recently, Bell's argument has been extended to show that realistic theories with arbitrarily high (but finite) influence propagation speed can neither reproduce the predictions of quantum mechanics \cite{hidden}. 

Inspired by Bell's work, Leggett proposed in 2003 a family of theories where the correlations generated in two-photon polarization measurements admit a very particular decomposition, that he termed \emph{crypto-nonlocal model} \cite{leggett}. As it turns out, certain correlations admitting a crypto-nonlocal model allow violations of local realism \cite{cyril}, and thus they were not ruled out by previous experiments of non-locality. Such experiments were conducted in \cite{nature,scarani, exp_it}.

Subsequent works on crypto-nonlocality have motivated/described Leggett's model by invoking the image that photons locally behave as if their polarizations were well-defined \cite{nature, scarani,cyril}. As a consequence, the general perception is that crypto-nonlocal theories satisfy the intuitive postulate that physical photon states have a definite polarization. We will call this axiom the \emph{realistic polarization principle}.

In this paper we show that the realistic polarization principle actually enforces constraints stronger than those captured by Leggett's crypto-nonlocal model. As a result, we find that the polarization measurement statistics of a multi-photon experiment in any no-signalling physical theory compatible with the realistic polarization principle must necessarily coincide with the correlations observed when measuring a fully separable state in quantum mechanics. Hence, all such theories are local realistic, and any quantum experiment verifying entanglement between the polarization degrees of freedom of $N>1$ photons can be considered a refutation of the realistic polarization principle. 

First, we will introduce the concept of crypto-nonlocality, as formulated by Leggett \cite{leggett}. We will then prove our result for the case of two photons and ideal measurements and discuss how to extend our arguments to more than two photons and non-ideal polarizers. Finally, we will present our conclusions.

Consider an experimental scenario where a source distributes pairs of photons to two parties, call them Alice and Bob. Alice (Bob) conducts a measurement of the polarization of her (his) photon in the direction $\vec{x}\in \C^2$ ($\vec{y}\in \C^2$) and outputs the value $a=0$ ($b=0$) if the photon is detected or $a=1$ ($b=1$), otherwise. Following Leggett \cite{leggett}, we say that Alice and Bob's statistics admits a \emph{crypto-nonlocal model} iff the probabilities $P(a,b|\vec{x},\vec{y})$ satisfy:

\be
P(a,b|\vec{x},\vec{y})=\sum_{\vec{u},\vec{v}}P(\vec{u}, \vec{v})P(a,b|\vec{x},\vec{y},\vec{u},\vec{v}),
\label{defin}
\ee

\noindent where $P(\vec{u}, \vec{v})$ is an arbitrary distribution of unitary vectors $\vec{u},\vec{v}\in \C^2$, and

\begin{eqnarray}
&P_A(a|\vec{x},\vec{u})\equiv\sum_bP(a,b|\vec{x},\vec{y},\vec{u},\vec{v})=\tr\{\Pi^{\vec{x}}_a\proj{\vec{u}}\},\nonumber\\
&P_B(b|\vec{y},\vec{v})\equiv\sum_aP(a,b|\vec{x},\vec{y},\vec{u},\vec{v})=\tr\{\Pi^{\vec{y}}_b\proj{\vec{v}}\},
\label{marginals}
\end{eqnarray}

\noindent with $\Pi^{\vec{z}}_c\equiv c\id_2+(-1)^c\frac{\proj{\vec{z}}}{\braket{\vec{z}}{\vec{z}}}$. Notice that $P_A(0|\vec{x},\vec{u})=|\hat{x}\cdot\vec{u}|^2$, $P_B(0|\vec{y},\vec{v})=|\hat{y}\cdot\vec{v}|^2$, i.e., locally, the subensembles satisfy Malus' law.

Remarkably, there exist non-local distributions compatible with these equations, see \cite{cyril} for a clarification of the relations between Leggett's crypto-nonlocality, Bell's nonlocality and quantum separability. Optimizing Bell-type functionals over all distributions admitting a crypto-nonlocal model is seemingly a very complicated mathematical problem \cite{scarani}.

Even though Leggett's model is \emph{inspired} from a number of physical considerations, like the reality of photon polarization, it has not been shown to be \emph{implied} by them. In \cite{nature}, it is nevertheless claimed that crypto-nonlocal theories ``are based on the following assumptions: (I) all measurement outcomes are determined by pre-existing properties of particles independent of the measurement (realism); (II) physical states are statistical mixtures of subensembles with definite polarization, where (III) polarization is defined such that expectation values taken for each subensemble obey Malus' law''. In the same line, it has been stated that ``the basic assumption of Leggett's model is that locally everything happens as if each single quantum system would always be in a pure state'' \cite{scarani}, or ``roughly speaking, it [the concept of crypto-nonlocality] says that all individual subsystems of a composite system should locally behave as if they were in a pure quantum state, with well-defined properties'' \cite{cyril}. 

Statements such as these reinforce the widely popular idea that Leggett's model arises from the intuition that physical photon states `should' have a well-defined polarization (or pure quantum state). In the following, we will show that photon polarization experiments in any reasonable physical theory where this realistic polarization principle holds cannot exhibit correlations beyond those obtained by measuring a separable quantum state. 

The class of physical theories that we will be considering satisfies the following axioms:

\begin{enumerate}[(a)]
\item 
Faster-than-light communication is impossible (no-signalling condition).
\label{nsc}
\item 
Physical photon states have a definite polarization (realistic polarization principle).
\label{rpp}
\item 
Given two separate photons, if we measure one of them and post-select on the result, the measurement statistics of the remaining system correspond to a photon state.
\label{crazy}
\end{enumerate}

\noindent Axiom (\ref{nsc}) is a consequence of relativistic causality \cite{sandu}, and axiom (\ref{rpp}) is a strong interpretation of axiom (II) in \cite{nature}, namely: ``physical states are statistical mixtures of subensembles with definite polarization'', see the discussion below. Axiom (\ref{crazy}) is a just re-statement of the intuition that post-selection is a type of preparation. To our best knowledge, this principle is satisfied in any physical model proposed so far, and it is implicit in the formalism of generalized probabilistic theories \cite{generalized1,generalized2,generalized3}. Furthermore, in our view, the concept of `two photons' encompasses this principle. We include it in the list of axioms in case that our notion of `two photons' is not shared by the reader.

Let us now investigate what kind of correlations should Alice and Bob expect to observe in bipartite photon polarization experiments should assumptions (\ref{nsc}), (\ref{rpp}), (\ref{crazy}) hold. W.l.o.g., suppose that Alice measures first, i.e., at time $t$ ($t+\epsilon$), Alice (Bob) carries out an action, consisting in either measuring her (his) system or not. Then axiom (\ref{rpp}) implies that Alice and Bob's correlations must be of the form (\ref{defin}), with the difference that, in principle, Alice's measurement can potentially influence Bob's physical state at a distance. That is, even though

\be
P(\emptyset,b|\emptyset,\vec{y},\vec{u},\vec{v})=\tr\{\Pi^{\vec{y}}_b\proj{\vec{v}}\},
\ee

\noindent if Alice chooses \emph{not} to measure her system at time $t$, it could be the case that $\sum_a P(a,b|\vec{x},\vec{y},\vec{u},\vec{v})\not=\tr\{\Pi^{\vec{y}}_b\proj{\vec{v}}\}$ if Alice conducts measurement $\vec{x}$ at that time. Due to axiom (\ref{nsc}), though, Bob's marginal statistics cannot depend on whether Alice measured her system or not (otherwise, Alice could signal Bob superluminally), and so we arrive at Leggett's model.

Note that, if we interpret axiom (II) in \cite{nature} in the sense that ``physical states are \emph{specific mixtures} of subensembles with definite polarization'' rather than \emph{arbitrary mixtures} (i.e., if, given a collection of subensembles $\{P(a,b|\vec{x},\vec{y},\vec{u},\vec{v})\}_{\vec{u},\vec{v}}$, we constrain the set of physical distributions $P(\vec{u},\vec{v})$ in eq. (\ref{defin})), then not all the subensembles $P(a,b|\vec{x},\vec{y},\vec{u},\vec{v})$ represent a physical state, and hence it is not clear why they should satisfy the no-signalling condition (\ref{marginals}). In principle, the no-signalling property of the physical state with distribution $P(a,b|\vec{x},\vec{y})$ could be recovered after averaging over $P(\vec{u},\vec{v})$, like in Bohm's theory \cite{bohm}, in which case the resulting model would reproduce the correlations generated by measurements of \emph{any} quantum state, separable or not.

We are not finished yet. Despite its apparent innocuity, axiom (\ref{crazy}) imposes extra constraints over eqs. (\ref{defin}), (\ref{marginals}). Indeed, imagine that Alice and Bob share the physical state with statistics $P(a,b|\vec{x},\vec{y},\vec{u},\vec{v})$, and Alice measures the polarization of her photon in the direction $\vec{x}\in \C^2$, obtaining the outcome $a$. By consecutive application of axioms (\ref{crazy}) and (\ref{rpp}), we have that Bob's resulting photon state is described by a convex combination of photon states with definite polarization. That is, there exist probability distributions $\mu^{\vec{x},\vec{u},\vec{v}}_a(\vec{w})$ over the unitary vectors $\vec{w}\in\C^2$ such that

\be
P(b|\vec{x},a,\vec{y},\vec{u},\vec{v})=\sum_{\vec{w}}\mu^{\vec{x},\vec{u},\vec{v}}_a(\vec{w})\tr\{\proj{\vec{w}}\Pi^{\vec{y}}_b\}.
\ee

\noindent This implies that

\begin{eqnarray}
&\tr(\proj{\vec{v}}\Pi^{\vec{y}}_b)=P(b|\vec{x},\vec{y},\vec{u},\vec{v})=\nonumber\\
&=\sum_a P(a|\vec{x},\vec{u},\vec{v})P(b|\vec{x},a,\vec{y},\vec{u},\vec{v})=\nonumber\\
&=\tr\left(\sum_aP(a|\vec{x},\vec{u},\vec{v})\rho^{\vec{x}}_a(\vec{u},\vec{v})\Pi^{\vec{y}}_b\right),
\label{paso1}
\end{eqnarray}

\noindent with $\rho^{\vec{x}}_a(\vec{u},\vec{v})\equiv\sum_{\vec{w}}\mu^{\vec{x},\vec{u},\vec{v}}_a(\vec{w})\proj{\vec{w}}$.

Since this relation holds for all unitary vectors $\vec{y}$, we have that

\be
\proj{\vec{v}}=\sum_a P(a|x,\vec{u},\vec{v})\rho^{\vec{x}}_a(\vec{u},\vec{v}).
\label{paso2}
\ee

\noindent Note that, in order to infer (\ref{paso2}) from (\ref{paso1}), it is sufficient that the considered physical theory allows conducting polarization measurements along three non-coplanar directions in the Bloch sphere (e.g.: two different linear polarization measurements plus circular polarization).

Now, $\proj{\vec{v}}$ is a pure state, so the above equation can only hold if $\rho^{\vec{x}}_a(\vec{u},\vec{v})=\proj{\vec{v}}$ for all $\vec{x}, a$. We thus conclude that

\begin{eqnarray}
&P(a,b|\vec{x},\vec{y},\vec{u},\vec{v})=P(a|\vec{x},\vec{u},\vec{v})P(b|\vec{x},a,\vec{y},\vec{u},\vec{v})=\nonumber\\
&=\tr(\proj{\vec{u}}\Pi^{\vec{x}}_a)\tr(\proj{\vec{v}}\Pi^{\vec{y}}_b).
\label{quid}
\end{eqnarray}

\noindent The statistics $\{P(a,b|\vec{x},\vec{y})\}$ hence arise from measurements of the separable state

\be
\sum_{\vec{u},\vec{v}}P(\vec{u},\vec{v})\proj{\vec{u}}\otimes \proj{\vec{v}}.
\ee

Conversely, it is easy to see that physical theories where only separable quantum polarization states can be prepared are compatible with axioms (\ref{nsc}), (\ref{rpp}), (\ref{crazy}). Such axioms are therefore equivalent to separable quantum mechanics, and so any two-qubit entanglement witness can be used to refute such a family of physical theories. 

This has to be contrasted with the results of \cite{cyril}, where it is shown that the correlations generated by certain two-qubit entangled states admit a crypto-nonlocal model. From our derivation it thus follows that such instances, while complying with Leggett's definition of crypto-nonlocality, cannot be present in any reasonable physical theory where the realistic polarization principle holds.

If we replace axiom (\ref{crazy}) by its obvious multiphoton extension (c'), our arguments can be generalized to more than two parties to conclude that only fully separable $N$-qubit states can be prepared in physical theories complying with axioms (a), (b), (c'). Indeed, suppose that $N$ parties share a multiphoton state with well-defined individual polarizations, and imagine that the first $N-1$ parties measure first, thus projecting the state of the $N^{th}$ party into a mixture of photons with definite polarizations. Again (\ref{paso2}) holds, and therefore, the last party is uncorrelated from the rest. The general result hence follows by induction.

Finally, accounting for imperfect measurements is easy: following Leggett's original paper \cite{leggett}, the response of non-ideal polarizers can be modeled by the measurement operator $\tilde{\Pi}_a^{\vec{x}}\equiv\epsilon'\id_2+(1-\epsilon)\Pi_a^{\vec{x}}$, with $0\leq \epsilon'\leq\epsilon<1$. Such operators are tomographically complete, and so, as in the ideal case, it is legitimate to infer eq. (\ref{paso2}) from eq. (\ref{paso1}). The rest of the proof is identical to the ideal case.

\vspace{10pt}
\noindent\emph{Conclusion}

We have shown that a reasonable interpretation of the intuition that ``polarization is well-defined'' leads, not to Leggett's definition of crypto-nonlocality, but to the strictly stronger notion of quantum separability. Hence, if the aim of the authors in \cite{nature} was to falsify the realism of photon polarization, there was no need to conduct any experiment whatsoever (polarization entanglement had been demonstrated long before \cite{entanglement}). Furthermore, in the absence of a physical motivation to propose crypto-nonlocality, we see no reason to carry out further research on this topic.

\noindent\emph{Acknowledgements}

We acknowledge interesting discussions with T. V\'ertesi, A. Ac\'in, M. Hoban, S. Popescu, C. Branciard and A. J. Leggett. This work has been supported by the John Templeton Foundation.

\end{document}